\documentstyle[prb,aps,epsf]{revtex}
\begin{document}
\title{
Transport properties of strongly correlated metals: a 
dynamical mean-field approach}
\draft
\author{Jaime Merino\cite{email0} and Ross H. McKenzie\cite{email} }
\address{School of Physics, University of New
South Wales, Sydney 2052, Australia}
\date{\today}
\maketitle
\widetext

\begin{abstract}
The temperature dependence of the transport
properties of the metallic phase of
a frustrated Hubbard model on the 
hypercubic lattice at half-filling are calculated.
Dynamical mean-field theory, which maps the
Hubbard model onto a single impurity Anderson model
that is solved self-consistently, and becomes exact
in the limit of large dimensionality, is used.
As the temperature increases there is a smooth
crossover from coherent Fermi liquid excitations at low
temperatures to incoherent excitations at high temperatures.
This crossover leads to a  non-monotonic temperature
dependence for the resistance, thermopower,
and Hall coefficient, unlike in conventional
metals. The resistance smoothly
increases from a quadratic temperature dependence
at low temperatures to large values which can exceed
the Mott-Ioffe-Regel value,
$ \hbar a/ e^2 $ 
(where $a$ is a lattice constant) associated with mean-free paths less than a
lattice constant.
Further signatures of the thermal destruction of
quasiparticle excitations are a peak in
the thermopower and the absence of a Drude peak 
in the optical conductivity.
The results presented here are relevant to
a wide range of strongly correlated metals,
including transition metal oxides, strontium
ruthenates, and organic metals.    
\\
\\
\end{abstract}


\section{Introduction}

The discovery of heavy fermion metals,
metal-insulator transitions in transition metal oxides,
high-temperature superconductivity in copper oxides,
and colossal magnetoresistance in manganates
has stimulated extensive theoretical studies
of strongly correlated electron models.\cite{fulde,gebhardt}
In spite of intensive research over the past decade
the nature of the metallic state in 
strongly correlated materials is still poorly understood.
This is particularly true of the 
cuprate superconductors, for which
most of the metallic properties
cannot be understood within the Fermi liquid picture
that has so successfully described conventional metals.\cite{liang}
Yet there are also a wide range of materials
which have low-temperature properties
(e.g., the observation of magnetic oscillations
such as the de Haas - van Alphen effect)
consistent with a Fermi liquid
but which at higher temperature are inconsistent with
a Fermi liquid.
These include transition metal oxides,\cite{imada}
 heavy fermions,\cite{stewart,cox,degiorgi}
strontium ruthenates,\cite{maeno}
the  quasi-two-dimensional
molecular crystals $\kappa$-(BEDT-TTF)$_2$X (Ref. \onlinecite{mck})
and the  quasi-one-dimensional Bechgaard salts\cite{jerome}
(TMTSF)$_{2}$X.
In conventional metals
the electronic properties are robust up
to temperatures of some sizable fraction of the Fermi energy.
In contrast, in the above materials
the electronic properties change at some
temperature much less than the Fermi energy.


A brief summary is now given of some
of the common differences between
the transport properties of strongly correlated
metals and the properties of elemental metals.
Later in the paper specific references will be
given to experimental results on a wide range of materials.

{\it Resistivity.}
Boltzmann transport theory gives
an expression for the magnitude of the resistivity
in terms of band parameters and a mean-free
path between quasi-particle collisions.
At low temperatures this expression
suggests a mean-free path which is much larger than
a lattice constant, as in conventional metals.
However, at higher temperatures the
resistivity smoothly increases to large values
that suggest a mean-free path which is smaller     than
a lattice constant, implying   the breakdown of
a quasi-particle picture.

{\it Thermopower.}
In conventional metals this is linear in temperature,
has values much less than 
 $k_B/e \simeq 87 \mu $ V/K, and has the same
sign as the charge carriers.
In strongly correlated metals it can have a non-monotonic
temperature dependence, can change sign, and
have values of the order $k_B/e$.

{\it Hall resistance.}
In conventional metals this is weakly temperature
dependent and gives the sign of the charge carriers.
In strongly correlated metals, the
Hall resistance can be strongly temperature dependent,
change sign, and  have the opposite sign
to the thermopower.

{\it Optical conductivity.}
In conventional metals, one observes a Drude peak
at zero frequency, which broadens but persists to high
temperatures. The spectral weight of this peak
is comparable to that predicted from the 
optical sum rule and the density of
charge carriers (or the plasma frequency).
 In contrast,
in strongly correlated metals most of the spectral weight 
is in broad features at high energies.
Furthermore, the Drude peak only exists at low temperatures.

\subsection{Dynamical mean-field theory}

The main purpose of this paper is to show
that transport properties such as those
described above are obtained in a dynamical
mean-field treatment of the Hubbard model.
Over the past decade a considerable amount of work has been
done using this approximation to understand
the Mott-Hubbard metal-insulator transition.\cite{vollhardt,Georges:96}
This approximation becomes exact in the limit of either
large  lattice connectivity or  spatial dimensionality.
It has been found to give a
good description of three-dimensional transition metal oxides
and has been argued to be relevant to the properties
of the cuprates.\cite{Pruschke:95,Georges:99}
Whereas most previous studies of transport properties 
\cite{Pruschke:95,Georges:99,Muller:99,palsson,Lange:99}
have focussed on doped Mott insulators
we consider the case where the band is half filled
and the Hubbard interaction $U$ is less than the
minimum value needed for the formation of the
Mott insulating state.
This is the situation in the metallic phase
of the 
molecular crystals $\kappa$-(BEDT-TTF)$_2$X (Ref. \onlinecite{mck}).

\begin{figure}
\centerline{\epsfxsize=12cm \epsfbox{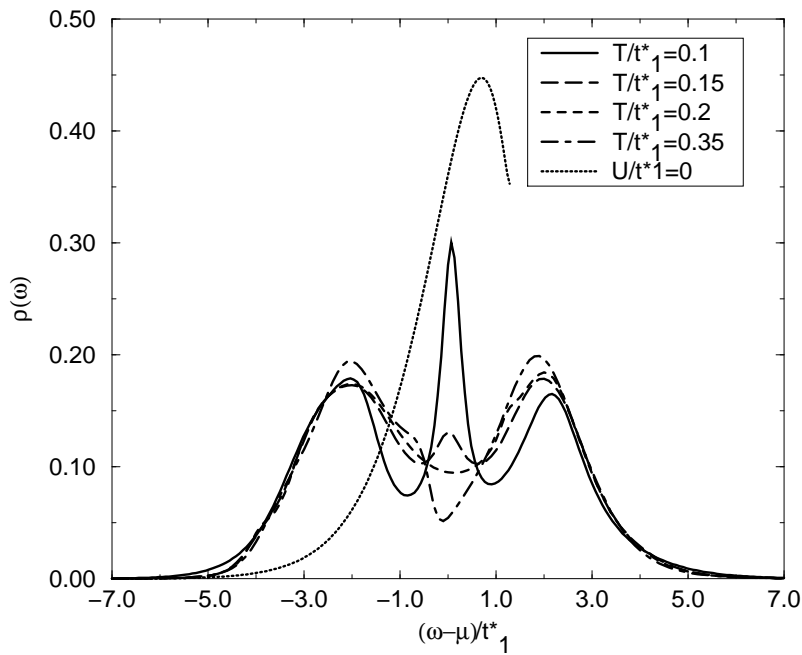}}
\caption{ Strong temperature dependence of the spectral density 
of the strongly correlated metallic phase of 
a Hubbard model at half-filling and
in large dimensions.       
Note that only at the low temperatures does 
a coherent quasiparticle band form near the
chemical potential, $\mu $.
The broad features near $\omega - \mu \simeq \pm U/2$
are the lower and upper Hubbard bands.
 The results 
shown are for $U = 4 t^*_1$ and a degree of 
magnetic frustration of $t_2^* = 0.3 t^*_1$. For comparison
we also show the non-interacting density of states ($U=0$), for which
the square root singularity placed at the upper band edge
is not plotted.
It is this strong temperature dependence of the spectral
density which
leads to many of the unconventional transport properties
discussed in this paper.}
\label{spectU4t203}
\end{figure}

Dynamical mean-field theory maps the Hubbard model onto a single impurity
Anderson model that must be solved self-consistently.
While time-dependent fluctuations are captured
by this approximation, spatially-dependent fluctuations
are neglected.     
Some important physics that emerges\cite{Georges:96} is that there is
a low-energy scale $T_0$ which is much smaller than
the non-interacting half-bandwidth $D$ and the Coulomb repulsion $U$.
$D$ is of the order of the 
 Fermi energy given by band structure calculations.
This energy scale  $T_0$ is the analogue of the Kondo
temperature for the impurity problem and defines the
energy scale of coherent spin excitations.
In the metallic phase the density of states $\rho(\omega)$
contains peaks at energies $\omega = -U/2$ and $+U/2$
 which correspond to the lower and upper
Hubbard bands, respectively, and involve incoherent
charge excitations.  These peaks
are broad and have width of order $D$.  At temperatures less than
$T_0$ a quasi-particle
peak with width of order $T_0$ forms at 
the Fermi energy (see Fig. \ref{spectU4t203} ).
The quasi-particle band  involves coherent excitations
(i.e., they have a well-defined dispersion relation)
that form a Fermi liquid.
The spectral weight of this peak (see Fig. 2) vanishes as the
metal-insulator transition is approached.
Thus, the temperature $T_0$ defines an energy scale
at which there is a crossover from Fermi liquid
behavior to incoherent excitations.
A similar crossover occurs in
 heavy fermion materials.\cite{stewart,cox,degiorgi}

\subsection{Overview}

In Section \ref{model} the model we study is introduced:
a Hubbard model on the hypercubic lattice with one
electron per site (i.e., at half-filling).
As well as a nearest-neighbour hopping $t_1$ a
next-nearest-neighbour hopping $t_2$ is also included
for several reasons. First, this term introduces
 magnetic frustration which enhances the
stability of the metallic phase
by suppressing 
the Neel temperature for antiferromagnetic ordering.\cite{Georges:96}
Second, in the absence of this term the model has
exact particle-hole symmetry and the
thermopower and Hall conductivity vanish.
Third, the model represents a higher-dimensional version
of a frustrated Hubbard model that describes the organic
conductors 
$\kappa$-(BEDT-TTF)$_2$X (Ref. \onlinecite{mck}).
In Section \ref{lisa} we review how the dynamical mean-field
theory reduces to an impurity problem.
In the infinite-dimensional limit all the vertex corrections
to correlation functions vanish and transport quantities are
determined by the one-electron spectral function.
The relevant expressions are presented in Section \ref{trans}.
Section \ref{ipt} describes how the local impurity
problem is solved at the level of iterated perturbation
theory. This method is known to give reliable results for
the impurity problem up to moderate interactions.

At low temperatures and low energies the electron self energy has 
a Fermi liquid form and in Section \ref{fl}
we present
analytical results for the different transport
quantities in this regime.
An expression is derived for the
Kadowaki-Woods ratio: the ratio of the $T^2$
coefficient of the
resistivity to the square of the linear specific heat coefficient
$\gamma$.
For strong interactions
it is shown to be independent of the band parameters and the
strength of the interactions.
The ratio of the thermopower to $\gamma T$ is shown to 
be independent of the strength of the interactions.

The temperature dependence of the different transport
quantities is presented in Section \ref{incoh}.
In particular we focus on the effect
 of the crossover from coherent to
incoherent excitations with increasing temperature.
For moderate to strong interactions the resistivity
smoothly increases from a $T^2$ dependence at low temperatures
to large values corresponding to mean-free paths less than a
lattice constant.
For strong interactions the resistivity can have a non-monotonic
temperature dependence; at temperatures several times
the coherence temperature it decreases with increasing temperature.
The thermopower is linear in temperature up to
a temperature of the order of the coherence temperature
at which it decreases. The resulting peak is similar
to the peak which occurs in the electronic specific heat and
is associated with the thermal destruction of the quasi-particles.
For strong interactions most of the spectral
weight in the optical conductivity is 
associated with transitions from (to) the lower (upper) Hubbard band.
A Drude peak only exists for temperatures less than the 
coherence temperature.

\section{Dynamical mean-field theory}
\label{sec:form}

\subsection{The Model}
\label{model}

We consider a Hubbard model  
with nearest neighbours hopping, $t_1$, and next-nearest 
neighbours hopping, $t_2$, on a given lattice.
The Hamiltonian is
\begin{equation}
H = t_1 \sum_{ij,\sigma} (c^\dagger_{i \sigma} c_{j \sigma} + h.c.)
 + t_2 \sum_{ik,\sigma} (c^\dagger_{i \sigma} c_{k \sigma} + h.c.) + 
U \sum_{i} n_{i\uparrow} n_{i\downarrow}- \mu \sum_{i \sigma} n_{i
\sigma} 
\end{equation}
where $U$ is the Coulomb repulsion between two electrons on
the same site and $\mu$ is the chemical potential.
We will only consider the case of half-filling, i.e.,
one electron per site.
 We treat the case of a $d$-dimensional
hypercubic lattice with connectivity $z$, which has a $t_1$ hopping to any of 
the $2z$ ($z=2d$) neighbours and $t_2$ along
the diagonals of the elementary unit cell.  
In order to have a finite kinetic energy in the $d \rightarrow \infty $ limit
the hoppings are rescaled as
$t^*_1 =\sqrt{2z}t_1$ and $t^*_2 =  \sqrt{2z (z-1)} t_2$,
with $z=2d$, $z$ being the connectivity of the lattice.
The non-interacting ($U=0$) density of states,
$D_0(\epsilon)=\sum_{k}\delta(\epsilon-\epsilon_k)$, 
associated with this lattice in the limit of infinite 
dimensions $(d \rightarrow \infty)$ reads \cite{Muller:89a,Georges:96}:  
\begin{equation}
D_0(\epsilon) = \left({2 \over \pi}\right)^{1/2}
 {{1}\over{E(\epsilon)}}
\cosh\left( { E(\epsilon) t^*_1 \over 2t^{*2}_2 }\right)
\exp \left({{t^{*2}_1-E(\epsilon)^2}\over{4t^{*2}_2}}\right)
\end{equation}
with $E(\epsilon)=[ t^{*2}_1+2t^{*2}_2-2 \sqrt{2} t^*_2 \epsilon ]^{1/2}$.
$D_0(\epsilon)=0$, whenever
 $E(\epsilon)$ is not real. Note that $D_0(\epsilon)$
has a finite band-edge with a square-root divergence. 
We set $t^*_1$ as the unit of energy.
The reason for choosing this lattice is that we can treat a 
varying degree of frustration by tuning the
ratio $t^*_2/t^*_1$,
which changes the shape of the bare density of states. Other
lattices, such as the Bethe lattice with next-nearest neighbours 
can also be used, but its density of states remains symmetric 
and therefore is qualitatively the same as its non-frustrated counterpart.

\subsection{Local Impurity Self-consistent Approximation (LISA) }
\label{lisa}

In the limit of infinite dimensions, mean-field theory   
of the full interacting lattice problem becomes exact 
and the problem reduces to  
solving a set of dynamical mean-field equations.\cite{vollhardt,Georges:96}
Therefore, the original Hubbard model is 
mapped to an impurity problem in the presence of a bath of electrons which
describes the rest of the lattice electrons and that has to be found 
self-consistently.  More precisely, one has to solve the associated
single-impurity Anderson model:
\begin{eqnarray}
H&=&\sum_{k,\sigma} (\epsilon_{k} - \mu ) c^+_{k\sigma} c_{k\sigma}
+\sum_{\sigma} (\epsilon_d-\mu) n_{d \sigma}
\nonumber \\
&+& \sum_{k,\sigma} 
V_{kd} (c^+_{d\sigma} c_{k \sigma} + c^+_{k \sigma} c_{d\sigma}) +
U n_{d\uparrow} n_{d\downarrow}
\label{HAnderson}
\end{eqnarray}
where the parameters $\epsilon_k$ and $V_{kd}$ describe the bath of electrons
through the hybridization function, which is defined as
\begin{equation}
\Delta(i \omega_n)=\sum_{k}{{|V_k|^2}\over{i \omega_n - \epsilon_k}}. 
\end{equation}
This function represents the amplitudes for the lattice electrons to
leave a site and, after wandering around the lattice, to return.
Therefore the problem remains local in space coordinates
but time-dependent correlations are fully taken into account. 
This is because in the large coordination 
limit, an electron can only hop once from one site 
to its nearest neighbour.
Processes, in which an electron can repeat a given path     
from one site to another in the lattice are suppressed
as they are at least of order $1/d$. 
Some preliminary work is just appearing \cite{Hettler}, which
tries to extend the zero-order expansion to include 
these type of higher-order processes. 

The bath function $\Delta(i \omega_n)$ is determined 
self-consistently, from the following condition: 
\begin{equation}
\Delta(i \omega_n)=i\omega_n - \Sigma(i\omega_n) - G^{-1}(i\omega_n)
\label{bath}
\end{equation}
where the self-energy $\Sigma(i\omega_n)$
is determined by solving the Anderson Hamiltonian
(\ref{HAnderson}), which is local in space,
 {\it i.e}, does not depend on 
momentum. $G(i \omega_n)$ is the lattice Green's function
from which the spectral densities can be obtained
\begin{equation}
\rho(\omega) =
-{{1}\over{\pi}}{ \rm Im} G( \omega+i \eta) 
\end{equation}

\subsection{Transport quantities}
\label{trans}

In the limit of infinite dimensions, transport quantities
can be calculated straightforwardly, due to the local nature
of the self-energy. For example, the evaluation of the optical 
conductivity simplifies drastically as only the particle-hole
bubble has to be evaluated in the Kubo formula. Contributions due
to higher order processes included in vertex corrections cancel
exactly\cite{Khurana}.
For a more detailed discussion of the derivation of
the expressions presented here see References
\onlinecite{Georges:96} and \onlinecite{Pruschke:95}.

Several transport    
quantities of interest can be obtained from
the spectral density.
The real part of the optical conductivity in the x-direction is given by
\begin{equation}
\sigma_{xx}(\nu)=\sigma_0
\int_{-\infty}^{\infty} d \omega 
{{ f(\omega) - f(\omega+\nu) } \over{ \nu} } {{1}\over{N} }
\sum_{{\bf k }\sigma}
 ( { {\partial \epsilon_{\bf k} }\over{ \partial k_x} } )^2
 \rho_{\bf k}(\omega) \rho_{\bf k}(\omega+\nu) 
\label{condxx}
\end{equation}
where $a$ is the lattice constant, $\sigma_0={{e^2 \pi}\over{
2 \hbar a} }$ and $N$ the total number 
of sites in the system.

The Hall conductivity is
\begin{equation}
\sigma_{xy}^H= \sigma_0^H \int_{-\infty}^{\infty}
d \omega { {\partial f(\omega)}\over{\partial \omega} }{{1}\over{N}} \sum_{\bf k \sigma}
({{\partial \epsilon_{\bf k}}\over{\partial k_x} })^2 {{\partial^2 \epsilon_{\bf k}}
\over{\partial k_y^2}} \rho_{\bf k}(\omega)^3
\label{condxy}
\end{equation}
where $B$ is an external magnetic field that points in the $z$
direction and $\sigma_0^H={{2\pi^2|e|^3aB}\over{3 \hbar^2} }$.  
We can also compute the 
Hall coefficient,  $R_H\equiv\sigma_{xy}/(\sigma_{xx}^2 B)$.

The thermopower is defined as
\begin{equation}
S=-{{k_B}\over{|e|T}} {{L_{12}}\over{L_{11}}} 
\label{thermo}
\end{equation}
where the transport integrals reduce in the $d \rightarrow \infty$ limit to:
\begin{equation}
L_{jk}=\int_{-\infty}^{\infty} d \omega \left( -  {{\partial f(\omega)}
\over{ \partial{\omega}} } \right) [ {{1}\over{N} }
\sum_{k \sigma} ( { {\partial \epsilon_k} \over{ \partial k_x} } )^2  
\rho_{k}(\omega)^2  ]^j \omega^{k-1}.
\label{transp}
\end{equation}

In the above expressions, a further simplification can be done
in the case of a simple hypercubic lattice, as all sums in momentum
reduce to integrations in energy weighed by the density of states.
\begin{eqnarray}
{{1}\over{N}}
\sum_{k \sigma} ( { {\partial \epsilon_{\bf k}} \over{ \partial k_x} }
)^2 \rho_{\bf k}(\omega)^2 &=& {{2}\over{d}} \int_{-\infty}^{\infty}
d \epsilon D_0(\epsilon) \rho(\epsilon,\omega)^2 
\nonumber \\
{{1}\over{N}} \sum_{\bf k
 \sigma}
( {{\partial \epsilon_{\bf k}}\over{\partial k_x} })^2 {{\partial^2
\epsilon_{\bf k}} \over{\partial k_y^2}} \rho_{\bf k}(\omega)^3
&=& -{{1}\over{2d^2}} \int_{-\infty}^{\infty} d\epsilon D_0(\epsilon)
\epsilon \rho(\epsilon, \omega)^3 \\
\end{eqnarray}
with the spectral densities given by
\begin{equation}
\rho(\epsilon, \omega) =      
-{{1}\over{\pi}}{\rm Im}
 \{ {{1}\over{\omega+\mu-\epsilon-\Sigma(\omega+i\eta)}} \}
\label{spectr}
\end{equation}
We will use this simplification in order to avoid the cumbersome
sums over momentum. In particular the dc conductivity reduces to
the following expression
\begin{equation}
\sigma_{xx}={{2 \sigma_0 }\over{d} }
\int_{-\infty}^{\infty} d \epsilon D_0(\epsilon )
\int_{-\infty}^{\infty} d \omega
\left( -{{\partial f(\omega)} \over{ \partial \omega} } \right)
\rho(\epsilon, \omega)^2 
\label{dc}
\end{equation}
for the simple hypercubic lattice.

For reasons of simplicity, we will still use the above
expressions in the presence of a non-zero $t^*_2$.
This is because the focus of this paper is on
many-body effects and not how different 
band structures may change the results slightly.

\subsection{Iterative Perturbation Theory}
\label{ipt}

A wide range of techniques have been used to solve the Anderson
 model (\ref{HAnderson}).
An extensive review has been given by Hewson \cite{Hewson}. 
Among them the iterative perturbation theory (IPT) is  
straightforward and at the same time gives a 
qualitatively correct description because it recovers exactly 
the atomic ($U/D \rightarrow \infty$) and the non-interacting 
($U=0$) limits.  It also provides
a  fast way for scanning a wide range of parameters in the Anderson
model, which, by means of other  methods such as 
exact diagonalization and quantum-Monte Carlo are
computationally very demanding.
Other approximate schemes such as the non-crossing 
approximation, which takes an infinite resummation 
of a certain class of perturbative diagrams, 
is limited in its applicability to high temperatures
\cite{Pruschke:93}.

Originally, the iterative perturbation scheme 
could only be applied for systems at half-filling   
and with particle-hole symmetry. This limitation
comes from the fact that the high energy behaviour of 
the spectral density is exactly reproduced for half-filling
by accident but this is not true
at arbitrary filling.
However, our main interest in this work is to study
frustrated models where the non-interacting density of states 
is non-symmetric and, consequently, particle-hole symmetry 
is broken. Recently
Kajueter and Kotliar\cite{Kajueter:96} have modified the standard
iterative perturbation theory to asymmetric cases, based on the 
earlier work of Martin-Rodero {\it et al} \cite{Martin:82}. 
Other authors\cite{Nolting:97}, have 
extended this framework to compute more accurately
the high-energy features of the spectral densities. Nevertheless, 
all of them approximately agree with exact 
diagonalization calculations when the interaction, $U$, is relatively
large. 

Our present work analyzes the transport
properties of metals which are 
strongly correlated but sufficiently away 
from the Mott transition so that well-defined quasiparticles 
exist at low temperatures. This means that we are always 
in the metallic side of the Mott transition but not too
close to the critical point at which the quasiparticle 
weight vanishes. Some controversy has arised recently
on the reliability of IPT even for moderate couplings of the interaction.
Very recently, Fischer \cite{Fischer} pointed out that a 
second-order expansion in terms of the 
skeleton diagrams which depend on the interacting Green's function,
$G(i \omega_n)$ instead of $G^0(i \omega_n)$, does not reproduce 
the upper and lower Hubbard bands: only a Fermi liquid type peak is
found in the spectral density\cite{Muller:89b}. The skeleton diagrams enter   
the expansion of the Luttinger-Ward functional and are the ones that
collapse into a local form in the $d \rightarrow \infty$, giving a local
self-energy \cite{vollhardt,Georges:96}. However, Yamada \cite{Yamada:76}
has shown that when
taking into account all the 4th order terms, the upper and lower 
Hubbard bands are reproduced, in agreement with the IPT results.
This means that an expansion up to second-order in the interacting
Green's functions is insufficient to grasp the correct behaviour of
the spectral density. Moreover, recent 
non-perturbative calculations done by Bulla,
Hewson, and Pruschke,\cite{Bulla:98}
using the numerical renormalization group  
for the Hubbard Model in infinite dimensions, clearly shows
the formation of the upper and lower Hubbard bands.            
Therefore, we believe that the method used here can be safely applied
giving a qualitative description of strongly correlated metals.
 
We use the finite-temperature version of the formalism, instead of the one
used by Kajueter and Kotliar\cite{Kajueter:96},
valid at $T=0$, as we are
interested in the thermodynamic properties of correlated
metals over a wide range of temperatures. 

We briefly outline the method used and refer the reader to
the more detailed work recently published
\cite{Kajueter:96,Nolting:99,Nolting:97}.

(i) Guess of an effective hybridization function: $\Delta(i \omega_n)$
and input of the chemical potential of the system $\mu$ together
with the chemical potential of the effective bath $\mu_0$. We 
fix the population per site of the interacting
lattice to be $n \equiv \langle n_\sigma \rangle=0.5$, and is 
kept fixed along the  rest of the steps.

(ii) Computation of the  Green's function of the effective 
bath:
\begin{equation}
G_0(i \omega_n)={{1}\over{i \omega_n + \mu_0 - \Delta(i\omega_n)}}
\end{equation}

and computation of the population of the bath: 
$ n_0 \equiv \langle n_{0\sigma} \rangle = -G_0( \tau=0^-)$

(iii) Ansatz for the self-energy, which is given by:

\begin{equation}
\Sigma(i \omega_n)= U n + {{A \Sigma^{(2)}(i \omega_n)}
\over{1-B \Sigma^{(2)}(i \omega_n )} }
\end{equation}
with $A$ and $B$ defined as 
\begin{equation}
A = {{n(1-n)}\over{n_0(1-n_0)}}  \ \ \ \ \
B = {{U(1-n)-\mu+\mu_0}\over{U^2 n_0(1-n_0)} } 
\label{AB}
\end{equation}
and the second-order self-energy
$\Sigma^{(2)}(i \omega_n)$ is computed from the imaginary
time-dependent Green's function of the bath 
\begin{equation}
\Sigma^{(2)} (i \omega_n )= \int_{0}^{\beta} d \tau e^{i \omega_n \tau} 
\Sigma( \tau ) 
\end{equation}
where $\Sigma(\tau)= -U^2 G^0(\tau)G^0(\tau)G^0(-\tau)$.  We 
use Fast Fourier Transforms to go back and forth from time to 
energy variables.  
The expression obtained for $A$, (\ref{AB}) comes from  
fixing the $m=2$ moment of the spectral density as 
explained in Ref. \onlinecite{Nolting:97}:
\begin{equation}
M^{(m)}= \int_{-\infty}^{\infty} w^m \rho(w) d w
\end{equation}
where $M^{(m)}$ can be computed from the Heisenberg equations of
motion. The parameter $B$ is fixed from the 
exact atomic limit solution, $V_{kd} \rightarrow 0$.  

(iv) Computation of impurity Green's function:

\begin{equation}
G(i \omega_n) = \sum_k G(i \omega_n, k)
= \int_{-\infty}^{\infty} {{D_0(\epsilon) d \epsilon} 
\over{i \omega_n + \mu - \epsilon -\Sigma(i \omega_n ) } }
\end{equation}

The free parameters $(\mu_0, \mu)$, can be now fixed from the
following set of equations
\begin{eqnarray}
n &=& -G(\tau=0^-)=0.5
\nonumber \\
n &=& n_0
\label{cond}
\end{eqnarray}
The last equation, originally proposed by Martin-Rodero
{\it et al.}\cite{Martin:82} is equivalent to the
Luttinger condition or the Friedel-Langreth sum rule and
fixes the correct low-energy behaviour of the self-energy.
Numerically this condition is much easier and faster to handle
than the Luttinger one.
Results from both of these conditions agree
 equally well
 with results from exact diagonalization of finite clusters.\cite{Nolting:97}
Finding $(\mu_0, \mu)$ takes around four to six iterations using Broyden's
method\cite{Fortran}.

(v) The final step is the requirement that the lattice Green's
funtion, $G(i\omega_n)$, coincides with the Green function of
the associated impurity problem given by the Anderson hamiltonian. 
This condition is expressed in equation(\ref{bath}).

The above steps (i)-(v) are repeated until a self-consistent bath
function is obtained. Note that the calculations are kept on  
the imaginary frequency axis: this makes the computation much faster and
more efficient with the use of Fast Fourier 
Transform algorithms. 
Analytical continuation to the real frequency axis
is needed  in order to 
compute the spectral densities entering the different transport
quantities. This continuation is numerically implemented
using Pad\'e approximants\cite{Vidberg:77}.

\section{Fermi liquid behavior at low temperatures}
\label{fl}

For temperatures and frequencies much less than
the Kondo temperature the self energy $\Sigma(\omega)$ of
the Anderson model has the
Fermi liquid form\cite{Hewson}
\begin{equation}
  \Sigma(\omega,T)= {\omega \over Z} - i C(\omega^2+(\pi k_B T)^2 )
\label{selfe}
\end{equation}
where $Z $ is the quasiparticle weight and
$C$ is a positive constant.
At sufficiently low temperatures and energies the
imaginary part becomes much less than the bandwidth and the
spectral function (\ref{spectr})
will have well-defined peaks when $\omega = Z E_k$
where $E_k$ is the band dispersion relation in
the absence of interactions.
The dependence of the quasiparticle weight $Z$
on the Hubbard interaction $U$ is shown in Fig. 2.
The specific heat  will be linear in temperature
at low temperatures with a slope that is
$1/Z$ times larger than the non-interacting value.
The effective mass $m^*$ deduced from magnetic
oscillations will also be larger than $m_b$, the value
predicted by band structure calculations, by the same factor
($m^*/m_b = 1/Z$).                              
This enhancement is found to be about two to four
in many organic metals,\cite{mck,wosnitza}
and Sr$_2$RuO$_4$\cite{mack0}.
In this section we consider the low-temperature
transport properties that follow from this form of the self energy.

\begin{figure}

\centerline{\epsfxsize=9cm \epsfbox{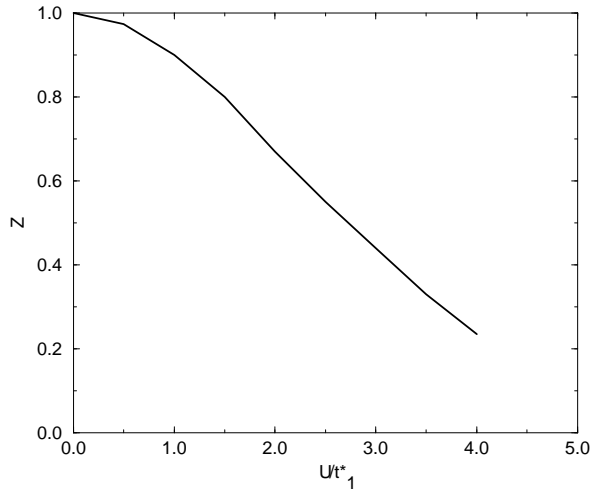}}
\caption{ Dependence  of the Fermi liquid quasiparticle weight, 
$Z$ on the Hubbard interaction $U$.
This paper focusses on the case of moderate 
interactions, $2 < U/t^*_1 < 4 $,
corresponding to effective mass enhancements
($m^*/m_b = 1/Z$) of two to four, as is observed
in many organic metals,\cite{mck,wosnitza}
and Sr$_2$RuO$_4$\cite{mack0}.
Even for such moderate interactions the
transport properties turn out to be strongly temperature
dependent. The curves shown are for $t^*_2=0.1 t^*_1$,
but virtually identical results are obtained for 
$t^*_2=0.3 t^*_1$.
}
\label{ZUt201}
\end{figure}

\subsection{Resistivity}
\label{sec:resis}      

The resistivity in a Fermi liquid behaves as 
\begin{equation}
\rho          \simeq A T^2
\label{rhoa}
\end{equation}
Such a temperature dependence is characteristic
of metals  in which the dominant scattering mechanism
is the interactions of the electrons with one
another and is observed in transition metals,\cite{imada} various
organic conductors\cite{mck}
and heavy fermions.\cite{stewart} 

Yamada and Yosida\cite{Yamada:86} demonstrated this
behaviour for an Anderson lattice and showed
that Umklapp scattering events should dominate the contribution
to the resistivity because momentum conservation 
would give an infinite conductivity when the lattice is not present.
Uhrig and Vollhardt have shown how in the limit of
large dimensions the umklapp processes lead to 
a finite conductivity.\cite{uhrig}
Cox and Grewe pointed out that in an anisotropic system
when the electron velocity and momentum are no
longer parallel that normal scattering can
contribute to the resistivity.\cite{cox}

In transition metals and heavy fermions
the Kadowaki-Woods rule\cite{kadowaki,woods}
relates the coefficient $A$
to the linear coefficient for the specific heat, $\gamma$:
$A/\gamma^2=$constant. The constant is
$4.0 \times 10^{-13} \Omega$cm (mol/mJ)$^2$
for transition metals, and
$1.0 \times 10^{-11} \Omega$cm (mol/mJ)$^2$
for most heavy fermions and
for transition metal oxides
near the Mott-Hubbard transition\cite{tokura}.
However, recent measurements on UPt$_{5-x}$Au found
values of 
$10^{-12} \Omega$cm (mol/mJ)$^2$
for $x=0, 0.5$ but it increases to 
$10^{-11} \Omega$cm (mol/mJ)$^2$
for $x > 1.1$.\cite{andraka}

We now evaluate the ratio $A/\gamma^2$ using our results.
From the self-energy (\ref{selfe}),
 the 
resistivity in the low-temperature limit associated 
with expression (\ref{dc}) is           
\begin{equation} 
\rho(T)\simeq {{2 d \sqrt{2\pi} k_B^2 \hbar a}\over{e^2 D I_{01}} } 
 C T^2
\label{rhot2}
\end{equation}
where we have numerically integrated
\begin{equation}
I_{nm} \equiv 
 \int_{-\infty}^{\infty}
{{ dx} x^n \over{(x^2 + \pi^2})^m} {{e^x}\over{(1+e^x)^2}} 
\label{Inm}
\end{equation}
and find $I_{01} \simeq 1/12$.
Expression (\ref{rhot2}) is the resistivity  at low temperatures for 
the case of a simple hypercubic lattice, for which the density of states
is $D_0(\epsilon) = {{1}\over{\sqrt{\pi}t^*_1}}e^{-\epsilon^2/t^{*2}_1 }$ and
$t^{*2}_1=4 t_1^2 d$.
$D$ is the effective half-bandwidth defined as $D=\sqrt{2} t^*_1$.

The linear specific heat term for the same density of states is
\begin{equation}
\gamma= {{2 \sqrt{2 \pi} \pi k_B^2}\over{3 Z  D} }
\label{gamma}
\end{equation}
where $Z$ is the quasiparticle weight.
Combining expressions (\ref{rhot2}) and (\ref{gamma})
we obtain
\begin{equation}
{{A}\over{\gamma^2}}={{9 d \sqrt{2 \pi} \hbar a}\over{4 \pi^3
k_B^2 I_{01} e^2 } } D C Z^2 .
\end{equation}
Hence, we see that if the dimensionless quantity
$D C Z^2$ is universal then so will
be the ratio $A/\gamma^2$. 
Insight into this question can be gained by considering first
a pure Anderson model, for which  we take a constant hybridization 
$\Delta=D$. For this case, it is found that\cite{Hewson}
$ C= {{(R-1)^2}\over{2 D Z^2} }$,
where $R$ is Wilson's ratio
\begin{equation}
R={\chi_{loc} / \chi_{loc}^0 \over \gamma / \gamma^0  }
\end{equation}
and $\chi_{loc}$ is the local susceptibility, $\gamma$ is
the linear coefficient for the specific heat, and the zero superscript
denotes the values in the absence of interactions.
$R$ takes values between 1 for $U=0$ and the universal 
value 2 for $|{{U}\over{D}}| \geq 1$
(Kondo regime).\cite{Hewson} 

We also find this scaling holds for the Anderson model with the
 self-consistent bath. 
We found $C$ by fitting the
imaginary part of the self-energy obtained from
our dynamical mean-field theory calculations
to the low-frequency and low-temperature
form (\ref{selfe}), for different values of $U$.  As shown in
 Table \ref{table1}, we find that 
$C$ scales with $1/Z^2$,  
for $U \geq 2 t_1^* $ as expected as we are in
the Kondo regime, giving a universal
behaviour of the $A/\gamma^2$ ratio. However,  
it decreases for $U \leq 2$ consistent with the result from the 
Anderson model that  $R \rightarrow 1 $ in the $U\rightarrow 0$
limit.

\newpage
\begin{table}
\caption{Values of the fitting parameter, $C$, and
quasiparticle weights for different values of the Coulomb repulsion, 
$U$ for the frustrated hypercubic lattice. Note that 
$C$ scales with $1/Z^2$ even for these
values of the interaction such that the metallic phase has well 
defined quasiparticles with only moderate enhancements
of the effective masses.       
}
\vskip0.1in
\label{table1}
\begin{tabular}{dddd}
$U/t^*_1$ & $C t^*_1$ & $Z(U)$ & $C Z^2t^*_1$ \\
\hline
1.  & 0.1 & 0.9 &  0.08  \\
1.5 & 0.23 & 0.8 &  0.16  \\
2.  & 0.44 & 0.67  &  0.20  \\
2.5 & 0.70 & 0.55 &  0.21 \\
3.  & 1.07  & 0.45 &  0.22  \\
\end{tabular}
\end{table}

From the numerical values of $C$ in Table \ref{table1},
we can compute the $A/\gamma^2$ ratio, using the
density of states of a simple hypercubic lattice: 
we get for $U \approx 3.0 $ (Kondo regime),
a ratio of $(1.24 a) \times 10^{-12} \Omega$ cm.
 This result is comparable to
experimental findings for  transition metal
oxides if we take the lattice constant to
be $a \approx 10 \AA$.

Previous calculations using a highly accurate
projective method to solve the dynamical
mean-field theory on the Bethe lattice
find that
very close to the Mott-Hubbard transition
$A/\gamma^2 = (2.3 a) \times 10^{-12} $ ohm cm (mol/mJ)$^2$
where $a$ is the lattice constant in $\AA$
of a three-dimensional system at half-filling.\cite{fisher}
This differs from our result by a factor of
two but turns out to be due to the different lattice used.
In order to make a direct comparison
with the results obtained in Reference \onlinecite{fisher}
we have repeated our calculations
using iterative perturbation theory
for a Bethe lattice at half-filling. 
We take a non-interacting density of states 
$D_0(\epsilon)=
{{1}\over{t_1^*\pi}} \sqrt{2-(\epsilon/t^*_1)^2}$.
The fitting parameters of the self-energy to
the low-temperature form for the Bethe lattice are shown in Table \ref{table2}.
We find that 
 already for moderate values of $U$, 
the value of $C Z^2$ converges rapidly to 
the value obtained 
in Reference \onlinecite{fisher}
providing a stringest test of the method used here.

\begin{table}
\caption{The same as in Table \ref{table1} for the Bethe
lattice. The final entry is
the result obtained in Reference \protect\onlinecite{fisher}
using a highly accurate projective method.
$U_c=4.13 t^*_1$ refers to the critical value at which 
the $T=0$ second order metal-insulator transition takes place.
}
\vskip0.1in
\label{table2}
\begin{tabular}{dddd}
$U/t^*_1$ & $C t^*_1$ & $Z(U)$ & $C Z^2t^*_1$ \\
\hline
1.5  & 0.35 & 0.72 &  0.18  \\
2.0 & 0.70 & 0.57 &  0.23  \\
2.5 & 1.40 & 0.42 &  0.25  \\
Uc=4.13 &- &- & 0.29 \\
\end{tabular}
\end{table}

\subsection{Thermopower}

Similarly to the above analysis for the resistivity     we 
can gain some insight into the behaviour of the thermopower
at low temperatures from the Anderson model.
It can be shown, either directly from Fermi
liquid theory or from large-N mean field theory \cite{Houghton}, that 
the thermopower increases linearly with temperature
at low temperatures.
Its slope scales as $1/Z$, in the same way as
the slope of the specific heat. Therefore, within the Anderson model 
the ratio of the thermopower
to the linear coefficient of the specific heat is independent of
Coulomb interaction. 
However, it depends on the band-filling: it drops to zero as
half-filling is reached, as it should, as for a system
with particle-hole symmetry and at half-filling the 
thermopower is zero.

 The  low temperature behaviour of the transport
integral, $L_{12}$, defined in (\ref{transp}), can be shown to be
\begin{equation}
L_{12}= \left.{{\partial D_0( \epsilon)}\over{ \partial \epsilon}} 
\right|_{\epsilon=\epsilon_F}
{{1}\over{2 \pi Z C} } I_{21}
\end{equation}
where $ I_{21}$ is the integral defined by (\ref{Inm}) and
the term proportional to the bare density states at the 
Fermi energy vanishes as the integral is antisymmetric.    
On the other hand $L_{11}$ is proportional to the DC 
conductivity and reduces for low temperatures  to the 
$T^2$ behaviour analyzed in subsection \ref{sec:resis}.
Therefore expression (\ref{thermo}) reduces to
\begin{equation}
S(T) = {{-k_B}\over{|e|}} \left. {  {{ \partial D_0( \epsilon)}\over{\partial \epsilon }}
\over{ D_0( \epsilon ) }  }  \right|_{\epsilon=\epsilon_F} {I_{21} \over I_{01}}{T\over Z} 
\label{thermop}
\end{equation}
where we, again, numerically compute the ratio of the 
integrals $I_{21}/I_{01} = 2.65$.
A similar expression was recently given by
Palsson and Kotliar, who considered the thermopower
in a doped Mott insulator.\cite{palsson}
The sign of the thermopower gives information on the
type of charge carriers (electron or holes) that are contributing
mostly to the transport. This
sign comes in our expressions from the slope of the 
density of states at the Fermi energy.

The ratio of the thermopower to the specific heat, at low
temperatures is given by
\begin{equation}
{{S(T)}\over{ \gamma T} } = -{{1}\over{|e|}}
{{3}\over{2 \pi^2}} \left.{  { {\partial D_0(\epsilon)}
\over{ \partial \epsilon } } \over{ D_0( \epsilon )^2 }  } 
\right|_{\epsilon= \epsilon_F }  {I_{21} \over I_{01}}  
\label{ratiother}  
\end{equation}
which is universal, i.e., independent of the interactions 
for a given degree of frustration in the lattice.

A simpler expression for the slope of the thermopower 
can be found in the limit, $t^*_2/t^*_1 \rightarrow 0$,
in this case, expression (\ref{thermo}) reduces to
\begin{equation}
S(T) \approx {{-k_B}\over{|e|}}  {I_{21} \over I_{01}}
\sqrt{2} {{t^*_2}\over{Z t^{*2}_1}} T
\label{thermopt2}
\end{equation}
The slope of the thermopower is, therefore, directly
proportional to the degree of frustration present in the 
frustrated hypercubic lattice. 
We have checked that at low temperatures our numerical
results are in good agreement with this expression. 

The simple expression (\ref{ratiother})
may explain the huge values ($S > k_B/e$ 
at 300 K)
recently observed\cite{terasaki}         
for NaCo$_2$O$_4$, 
which has potential applications
as a thermoelectric material.\cite{mahan2}
This material consists of layers of CoO$_2$
with the crystal structure
of a  triangular lattice. For such a lattice the 
non-interacting density of states can be
expressed analytically as shown in Ref.
\onlinecite{Ivanov}. Evaluating
the derivative at the Fermi energy for a half-filled band
gives 
\begin{equation}
 \left.{  { {\partial D_0(\epsilon)}
\over{ \partial \epsilon } } \over{ D_0( \epsilon )^2 }  } 
\right|_{\epsilon= \epsilon_F }  = -1.24
\label{triang}  
\end{equation}
and so (\ref{ratiother}) predicts a ratio of $1/2|e|$,
which re-expressed in appropiate units 
is: $5.23 \times 10^{-3} \mu$ V mol/ mJ. The measured thermopower 
is approximately linear in temperature up to
about  $T=$ 200 K, at which it has a value of
 about        $ 80 \mu$ V/K.
The measured specific heat coefficient\cite{terasaki2} is
$\gamma = 48 $ mJ/ (mol K$^2$) giving a ratio $(S(T)/\gamma T)$
is $8 \times 10^{-3} \mu$ V mol/ mJ.
This suggests that the large value of the thermopower
of this material is not just due to strong correlations
enhancing the effective mass
but also due to the large particle-hole asymmetry
associated with the triangular lattice.
Also, the theory presented predicts a positive
thermopower at low temperatures for the triangular
lattice, consistent with experiment.\cite{terasaki}

\subsection{Hall resistance}

In the low-temperature limit,  the Hall conductivity (\ref{condxy})
reduces to 
\begin{equation}
\sigma^H_{xy} = {\sigma^H_0 \over{2 d^2} }  {{3}\over{8 \pi^2}}
D_0(\epsilon=\epsilon_F) {{\epsilon_F }\over{C^2 T^4}}
I_{02}
\label{HallTlow}
\end{equation}
where the integral $I_{02}$ is defined by (\ref{Inm}) and
is equal to 0.00730.
This expression
depends on the interaction through, $C \propto 1/Z^2$. 
A similar expression was recently found by
 Lange and Kotliar.\cite{Lange:99} 

The Hall coefficient 
reduces at low temperatures to 
\begin{equation}
R_H \approx {{a^3}\over{6 |e|}} {{I_{02}}\over{I_{01}^2}}
{ {{\rm Re} \Sigma(\omega =\mu) - \mu}\over{ t^{*2}_1 D_0(\epsilon_F)} }
\label{RH}
\end{equation}
where $I_{02}/I_{01}^2=1.06$. Expression (\ref{RH})  shows  
temperature dependence through the chemical 
potential $\mu=\mu(T)$.  At $T=0$, expression (\ref{RH})
is independent of $U$ because                      
from Luttinger's theorem  
$ \epsilon_F \equiv {\rm Re} \Sigma(\omega =\mu) - \mu $. 
Moreover, in the $U \rightarrow 0$ limit, and in the particle-hole symmetric case,   
expression (\ref{RH}) reduces to zero, $R_H = 0$ for $T \rightarrow 0$,
as it should, as the density of holes cancels exactly the 
density of electrons contributing to the transport in
the system.  
As soon as the degree of frustration $t_2^*/t_1^*
\neq 0$, then, the Hall coefficient is non-zero, and, again 
at $T=0$, independent of the Coulomb interaction.
The sign of the Hall coefficient depends on the sign of        
the real part of the self-energy referred to the chemical
potential. This means that, in general, it is possible to 
have a different sign for
the thermopower and the Hall factor at low temperatures 
depending on the shape of the bare density of states and 
the Fermi energy.

\subsection{Optical conductivity}

At low temperatures the
optical 
conductivity (\ref{condxx}) reduces to                    
\begin{equation}
\sigma(\omega) = {\sigma_0 D(\epsilon_F) \over d \pi }
\int d \nu { f(\nu)-f(\nu+\omega) \over \omega } 
{( 1/\tau(\nu)+ 1/\tau(\nu+\omega) ) \over ( Z \omega )^2
+ {1\over 4} ( 1/\tau(\nu) +  1/\tau(\nu +\omega) )^2  }
\end{equation}
where $1/\tau(\nu) = 2 {\rm Im} \Sigma(\nu)$, similar to an
expression  
first obtained by Murata\cite{Murata}.
For $\omega <<\pi T $,
 the frequency dependence of the self-energy can be neglected
and the above expression 
 reduces to
\begin{equation}
\sigma(\omega,T) = {2 \sigma_0 D_0(\epsilon_F) Z \over  d \pi }  
 {\tau^*(T) \over 1 + (\omega \tau^*(T))^2 }
\label{drude}
\end{equation}
where $1/\tau(T) = 2 {\rm Im} \Sigma(0,T) \approx 2 C (\pi T)^2$ and
$\tau^*(T)=\tau(T)/Z$.

\section{Crossover to incoherent excitations}
\label{incoh}

The Fermi liquid behaviour discussed in the previous
section only occurs up to some temperature of the
order of the coherence temperature $T_0$.
There is then a smooth crossover to the case where
all of the low-energy excitations are incoherent (see Fig. 1).
In this section we present results showing the
effect of this crossover on transport properties.

\subsection{Resistivity}

\begin{figure}
\centerline{\epsfxsize=9cm \epsfbox{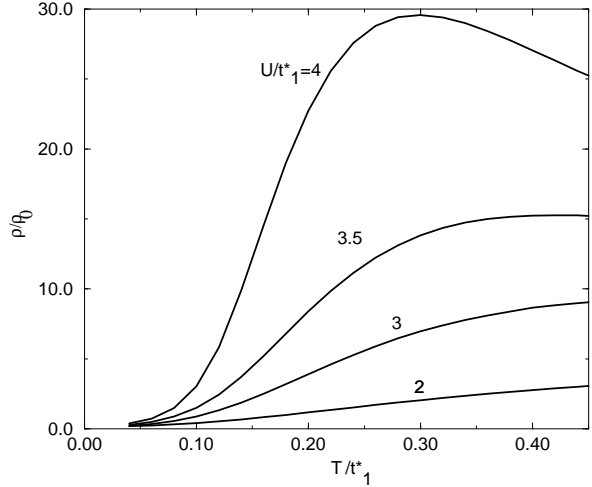}}
\caption{ Temperature dependence of the resistivity in the
frustrated hypercubic lattice for $t^*_2/t^*_1=0.1$ and
for different values of $U/t^*_1=2, 3, 3.5$ and 4. 
For $U=4 t^*_1$ there is a crossover from metallic behaviour at 
low temperatures to insulating                        
behavior at high temperatures.
The resistivity is given in units of $\rho_0 = 
\hbar a /e^2$, where $a$ is a lattice constant
which corresponds to a value of the resistivity at
which the mean free path is comparable to a lattice constant.
\label{resistt201}}
\end{figure}

The temperature dependence of the resistivity is shown in
Fig. \ref{resistt201}  for $t^*_2/t^*_1= 0.1$
and various interaction strengths. 
It has two properties often observed
in strongly correlated metals:
(i) for strong interactions a non-monotonic
temperature dependence occurs, and
(ii) for high temperatures it smoothly
increases to large values corresponding
to mean-free paths less than a lattice constant.

For values of the interaction comparable to the bandwidth,
$U \approx 4 t_1^*$,
the resistivity shows a peak at a temperature of about
$0.2 t^*_1$.
The temperature at which this peak appears
corresponds approximately to  the temperature at which
there are no longer Fermi liquid quasiparticles
present (see Fig. 1).
The decreasing resistance with increasing temperature,
characteristic of  a semiconductor or insulator,
is due to thermal excitations to the upper Hubbard band.
Note that the peak       temperature is not the Kondo temperature,
which, in our calculations is at much lower temperatures.
Such a 
peak in the resistivity is observed                   
in heavy fermion sytems\cite{stewart}and 
some of the 
$\kappa-$(BEDT-TTF)$_2$X
family of
organic superconductors\cite{mck}.
Results similar to Fig. \ref{resistt201} were obtained previously
 for the simple hypercubic lattice ($t_2 =0$)
when the impurity problem was solved using
Quantum Monte Carlo and the non-crossing approximation,\cite{Pruschke:93}
and for the Bethe lattice when the impurity
problem was solved using iterative perturbation theory.\cite{roz2}

{\it Bad metals.}
In conventional metals transport occurs
by well-defined quasi-particles; they  have a
wavelength ($\sim 1/k_F$)
much less than the mean-free path $\ell$
and so transport properties can be described 
by the Boltzmann  equation.
However, if the scattering is sufficiently strong
that the mean-free path is comparable to
a lattice constant ($\ell \sim a$) then
$k_F \ell \sim \pi$ and the quasi-particle concept
breaks down.
This is often referred to as the Mott-Ioffe-Regel limit.\cite{mott}
For an isotropic three-dimensional metal this corresponds
to a conductivity of $\sigma = e^2 /(3 \hbar a),$
and is sometimes referred to as the Mott minimum
conductivity.
However, for a wide range of strongly correlated metals,
including the cuprates\cite{gurvitch}, fullerene
metals (A$_3$C$_{60}$) (Ref. \onlinecite{hebard}), the organic
superconductors $\kappa-$(BEDT-TTF)$_2$X (Ref. \onlinecite{mck}),
Sr$_2$RuO$_4$, (Ref. \onlinecite{tyler}),
SrRuO$_3$ (Ref. \onlinecite{klein}), and VO$_2$ (Ref. \onlinecite{allen}),
it is observed that as the temperature
increases the resistivity can increase to values corresponding
to mean-free paths much less than a lattice constant.
Such materials have been referred to as
``bad metals.''\cite{emery}
In contrast, in the A-15 metals
the resistivity appears to `saturate' at a high temperature
value corresponding to the Mott-Ioffe-Regel limit.\cite{fisk}
However, it has recently been suggested that the
resistivity does not saturate but rather a change
in temperature dependence occurs when the
scattering is strong enough to cause a breakdown of the
Migdal approximation.\cite{millis}
Emery and Kivelson proposed\cite{emery} that the
smooth temperature dependence of the
resistance in bad metals suggests that
the low-temperature transport is also not due
to quasi-particles.

At low temperatures the resistivity given by (\ref{rhot2})
can be written 
\begin{equation}
\rho         \simeq  d (2 \pi)^{1/2} { \hbar a \over e^2 }  
{1  \over \tau D}.
\label{mottr}
\end{equation}
where $\tau$ is the scattering time.
The Mott-Ioffe-Regel condition ($\ell \simeq a$)
is equivalent to
$\tau  D \simeq 2 \pi$, leading to
a resistivity
\begin{equation}
\rho_{0} \simeq  {\hbar a \over e^2 }.
\label{mott2}
\end{equation}
For $a=10 \AA$ this corresponds to a resistivity of  3 m$\Omega$-cm.
Fig. \ref{resistt201} shows that even for moderate
interaction strengths 
the resistivity can smoothly increase to
values much larger than this. 
Furthermore, our
results provide a counter example to the
 ideas of Emery and Kivelson\cite{emery}
since there is a smooth crossover from transport by incoherent excitations
at high temperatures 
to  Fermi liquid transport at low temperatures.

\subsection{Thermopower}

\begin{figure}
\centerline{\epsfxsize=9cm \epsfbox{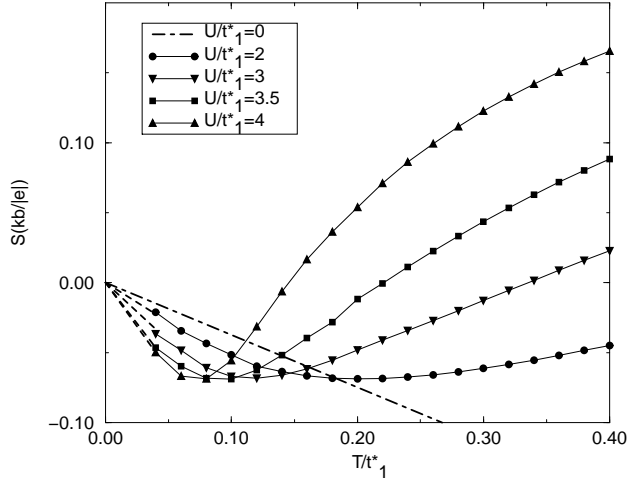}}
\caption{ Temperature dependence of the thermopower 
for the
frustrated hypercubic lattice for $t^*_2/t^*_1=0.1$ and
for different values of $U/t^*_1=2, 3, 3.5$ and 4.
The dashed lines are based on linear extrapolations
to zero temperature
as expressed in Eq.(\ref{thermopt2}). The curves show how 
interactions between electrons 
significantly change the magnitude and temperature
dependence from the
linear behaviour expected for a 
weakly interacting Fermi liquid.
Indeed,
the appearance of a minimum in the thermopower 
is a signature of thermal destruction of the
coherent Fermi-liquid state which exists at low temperatures.
\label{thermot201}}
\end{figure}

 In Fig. \ref{thermot201}, we show the 
results for the thermopower as a function of temperature
for different values of the Coulomb interaction, $U$ and
in the nearly symmetric case $t^*_2/t^*_1=0.1$.

\begin{figure}
\centerline{\epsfxsize=9cm \epsfbox{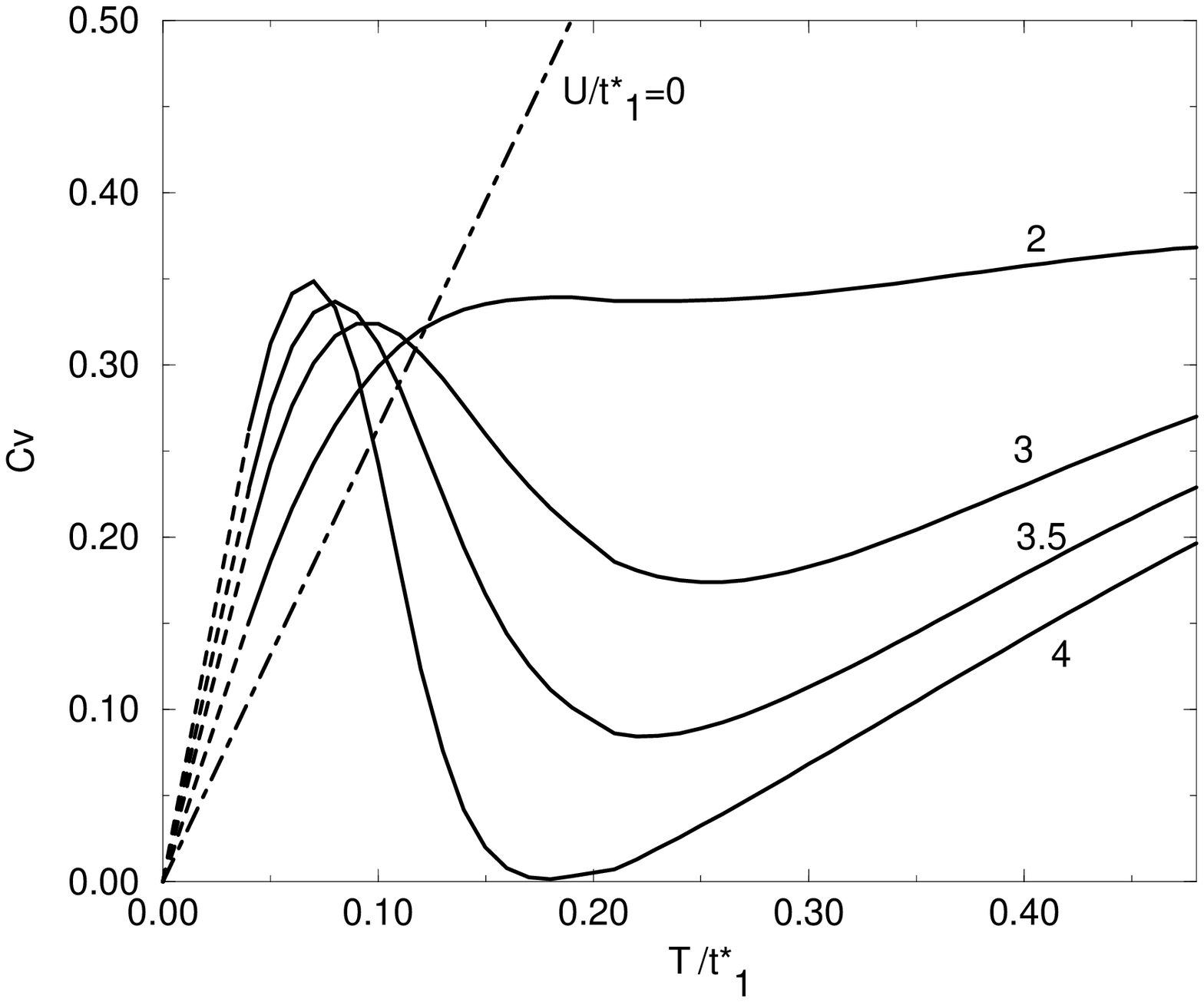}}
\caption{ Specific heat for 
the frustrated hypercubic lattice for $t^*_2/t^*_1=0.1$ and
for different values of $U/t^*_1=2, 3, 3.5$ and 4. Note
that the peak occurs at a temperature comparable to that 
at which the peak in the thermopower occurs (compare 
with Fig. \ref{thermot201}). The dashed lines are 
linear extrapolations to zero temperature. As in the case
of the thermopower, the simple linear behaviour found
for the non-interacting case, is qualitatively 
changed and a peak at low temperatures shows up for
$U/t^*_1 \geq 3$. The temperature scale at which this peak appears
is set by the binding energy of the 
spin-screening cloud formed at each lattice site 
due to the Kondo effect.}
\label{Cespt201}
\end{figure}

The low temperature behaviour is 
correctly described by equation (\ref{thermo}). 
As it can be observed, the 
slope of the thermopower at low temperatures   
increases with increasing $U$, scaling as 
the effective mass $m^*/m=1/Z$. 
We find a minimum in the thermopower 
which is rather shallow for small $U$ and becomes
increasingly pronounced with increasing $U$. A similar feature
also occurs for doped Mott insulators\cite{Pruschke:95} 
and for the Anderson lattice.\cite{Czycholl:91}
We observe that the mimimum moves to 
higher temperatures as $U/t^*_1$ is decreased. This is 
a consequence of the increase in the Kondo scale
with decreasing $U $ and is supported by      
the observation that this minimum follows the
peak in the specific heat. 
To illustrate the close relationship between the
thermopower and the specific heat,
Fig. \ref{Cespt201} shows the specific heat 
for the same parameter values as Fig.       
\ref{thermot201}. 

The peak
in the specific heat, already analyzed by several authors
\cite{Krauth:93}, is associated with the binding energy of the 
Kondo spin screening cloud which eventually forms at each lattice site.        
The high temperature behaviour found is typical of 
systems which have a depleted density of states at the Fermi energy,
for example, in semimetals and insulators one expects the 
magnitude of the thermopower to decrease as the temperature is 
decreased. This is more easily understood from the behaviour of the
spectral densities which show
this effective depletion of quasiparticle excitations at the
Fermi energy (see Fig.\ref{spectU4t203}). 

The change in sign of the thermopower at 
intermediate temperatures $T \approx 0.2 t^*_1$ for $U = 3.5 t^*_1$,
can be explained from the fact that the spectral weight 
of the quasiparticle excitations is transferred mostly to the lower 
rather than to the upper  Hubbard band,
making the holes, rather than  the electrons,  
the dominant carriers contributing to energy
transport (see Fig.\ref{spectU4t203}). 

It is worth stressing that it is not necessary to
get to too large values of $U/t^*_1$ to find a 
clear signature of the minimum in the thermopower and 
strong temperature behaviour.         
This is a feature which one can find in sufficiently 
correlated systems far from the Mott transition as
can be checked from the effective masses we obtain
, $m^*/m$, which in our
calculations vary between 2-4 for $ U \approx 2$
and 4, respectively. 
  
\begin{figure}
\centerline{\epsfxsize=9cm \epsfbox{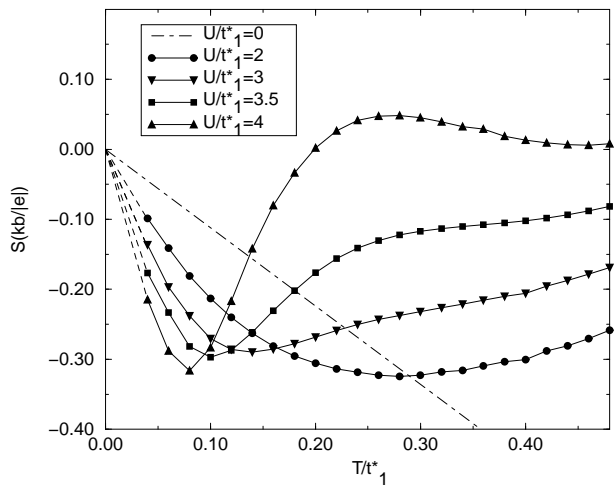}}
\caption{ Temperature dependence of the thermopower for the
frustrated hypercubic lattice with $t^*_2/t^*_1=0.3$ and 
$U/t^*_1=2, 3, 3.5,$ and 4.
The dashed lines are extrapolations to zero temperature. 
The effect of frustration is more clearly seen when
comparing this figure with the case $t^*_2/t^*_1=0.1$. The values of
the thermopower are increased as well as the slope at low
temperatures. However, the position of the minimum is 
nearly independent of the degree of frustration.
\label{thermot203} } 
\end{figure}

Fig.\ref{thermot203} shows the thermopower when the frustration is 
increased to $t^*_2/t^*_1 =0.3$. The magnitude of the
thermopower is enhanced as a result of the larger
asymmetry present in the particle-hole excitations
of the system. Thus, the slope at low temperatures
is increased by a factor of about   3,   as expected
from Eq.(\ref{thermopt2}). The main
features remain similar to the less frustrated
case $t^*_2/t^*_1 \approx 0.1$, although the minimum of the
thermopower is more pronounced for a larger degree of
frustration.

The thermopower of
the cuprates\cite{zhou} and the organic superconductors
$\kappa$-(BEDT-TTF)$_2$X (Ref. \onlinecite{yu}),
and
$\beta$-(BEDT-TTF)$_2$X (Ref. \onlinecite{beta2})
have the common
properties that it is not a monotonic function
of temperature and has large values of order
10-50 $\mu$ V/K at 100 K.
For the organics these properties cannot be
explained in terms of the calculated band structures and
a weakly interacting Fermi liquid.\cite{yu,beta2}
For 
Sr$_2$RuO$_4$
the thermopower increases nonlinearly with
temperature from 4 K to 300 K, appearing
to saturate at high temperatures,
and has the opposite sign as the Hall coefficient.\cite{yoshino}

As discussed above a peak or minimum in 
the thermopower is a signature               
of the decay of coherent excitations with increasing temperature.
It is desirable to see if this feature
can be observed in experiments on
other strongly  correlated metals.
Such a peak should be clearly distinguishable
from a peak due to phonon drag\cite{barnard} by several features.
The latter produces a thermopower which is
proportional to the lattice specific heat and thus
cubic in the temperature for $T \ll \theta_D$.
For higher temperatures the phonon drag
thermopower goes like $1/T$.
The result is a peak around a temperature
of 0.1-0.2 $\theta_D$.
Values of $\theta_D$ can be deduced
from the specific heat data.
Thus, it should be possible to distinguish whether
an observed peak in a material is due to
phonon drag or due to loss of Fermi liquid coherence
because of the different temperatures at which they
occur and because of the different behaviour at
higher temperatures.

Zhou and Goodenough
have observed peaks around 100 K in the thermopower of
CaVO$_3$ \cite{zhou2} and
La$_{1-x}$Nd$_x$CuO$_3$.\cite{zhou3}
They attribute these peaks to phonon drag.
This peak  cannot be due to the correlation effects considered here
because it occurs at too low a temperature.
In CaVO$_3$ the optical conductivity
still has a Drude peak at 300 K,\cite{roz}
and it is estimated that
$D$ = 1 eV and $U$ = 3 eV.
Consequently, the coherence temperature will be of the order 
of 1000 K.

The peak in the electronic specific heat would
be extremely difficult to observe because it will be
masked by the $T^3$ phonon contribution.
In contrast, the phonon contribution to the thermopower
decreases with increasing temperature and so should not
mask the feature due to correlations.

\subsection{Hall resistance}

\begin{figure}
\centerline{\epsfxsize=9cm \epsfbox{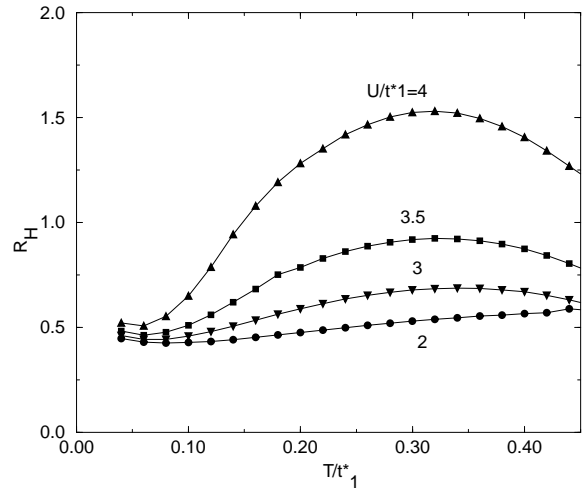}}
\caption{ Temperature dependence
of the Hall coefficient for the frustrated hypercubic
lattice with $t^*_2/t^*_1= 0.1$  and
for different values of $U/t^*_1=2, 3, 3.5$ and 4.
Note the saturation of the Hall coefficient at low temperatures
for all values of $U/t^*_1$. As the interaction is increased
strong temperature dependence arises.   
\label{Hallt201}}
\end{figure}

\begin{figure}
\centerline{\epsfxsize=9cm \epsfbox{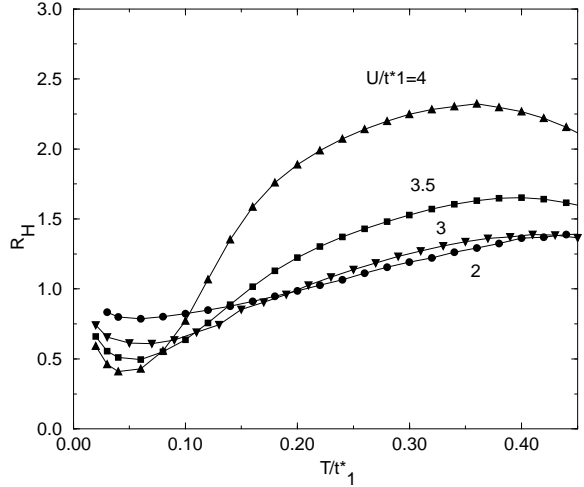}}
\caption{ Temperature dependence of the
 Hall coefficient for the frustrated hypercubic 
lattice with $t^*_2/t^*_1= 0.3$  and    
for different values of $U/t^*_1=2, 3, 3.5$ and 4.
A greater degree of assymetry can enhance some features
of the Hall resistance. At low temperatures, it is more strongly
temperature dependent than for $t_2^*/t_1^*=0.1$ and even an
upturn in the Hall coefficient can arise.
\label{Hallt203}}
\end{figure}

In Fig.\ref{Hallt201} and \ref{Hallt203}
we show results for the 
temperature dependence of the Hall coefficient $R_H$ for
$t^*_2/t^*_1=0.1$ and 0.3, respectively. 
We observe from the curves that 
for small values of the $U/t^*_1$ ratio, the Hall coefficient
is nearly independent of temperature, whereas for larger values of 
the interaction there is an increase in the Hall coefficient
for increasing $T$ reaching a maximum at $T \approx 0.3 t^*_1$.
This fact is observed for both    values of the frustration shown.
Note that the sign of the Hall coefficient is not
necessarily the same as the sign of the thermopower.

For a given value of the frustration,
all curves converge to the same value at $T=0$ as expected from 
Eq.(\ref{RH}).
However, the temperature dependence at low
temperatures for $t_2^*/t_1^*=0.3$ differs from the 
$t_2^*/t_1^*=0.1$ case. The behaviour at low temperatures
is determined by the temperature dependence of the 
chemical potential, which depends on the lattice analyzed through 
the bare density of states and the value of $U/t^*_1$.
For the case $t_2^*/t_1^*=0.3$, the Hall coefficient is more strongly 
dependent of $U$ than for
the $t^*_2/t^*_1=0.1$ case. Moreover, an upturn of the Hall coefficient
is found in the low-temperature limit
 $T \rightarrow 0$ in the latter case. This means that, although
qualitatively the situation is similar for different degrees
of frustration, some features can be enhanced and may depend
on the details of the band structure and the bare density of states
of the material. 

The Hall coefficient for a doped Mott insulator
 on a simple hypercubic lattice was calculated previously by Pruschke
{\it et al.}\cite{Pruschke:95} and Lange and Kotliar\cite{Lange:99}
using dynamical mean-field theory and found
to have a qualitatively similar temperature dependence.

The layered perovskite Sr$_2$RuO$_4$ has Fermi liquid
properties at low temperatures\cite{maeno} but the
Hall resistance of Sr$_2$RuO$_4$ is strongly
temperature dependent.\cite{mack}
It has a value of about -1.15 $\times 10^{-10}$ m$^3$ C$^{-1}$
below 1 K and then increases rapidly with temperature
and changes sign around 35 K and
saturates at high temperatures to a value
of about -0.1 $\times 10^{-10}$ m$^3$ C$^{-1}$.
The behaviour and value below 1 K can
be explained within a Fermi liquid picture.\cite{mack}
However, the sign change can only be explained
if the temperature dependence of the scattering
rate in the different bands is significantly different.\cite{mazin}
An alternative explanation for the temperature
dependence is the decay of coherence discussed here.

Experiments on organic metals 
 $\kappa$-(BEDT-TTF)$_2$X               
show a temperature-dependent Hall coefficient.\cite{sushko}
For $\beta$-(BEDT-TTF)$_2$I$_3$
the Hall resistance has a broad maximum around 40 K.\cite{cooper}

\subsection{Optical conductivity}

\begin{figure}
\centerline{\epsfxsize=9cm \epsfbox{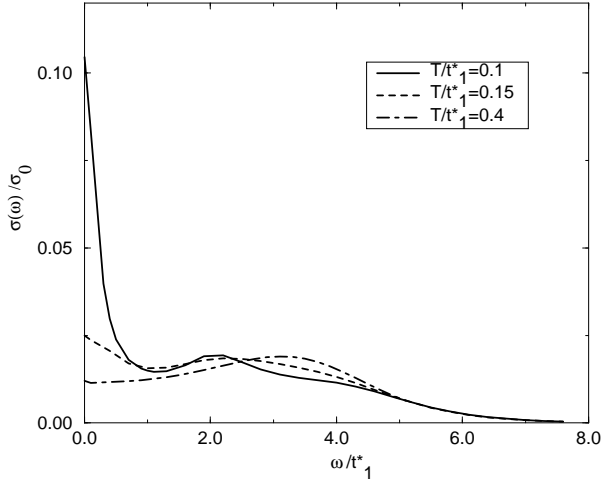}}
\caption{ Strong temperature dependence of the 
optical conductivity. The curves shown
are  for $U = 4 t^*_1$,
$t^*_2= 0.1 t^*_1$, at three different temperatures.
A Drude peak at zero frequency only occurs at low temperatures.         
The feature around
$\omega \approx U/2 $ is due to transitions from the coherent
quasiparticle band at the Fermi energy to the upper Hubbard band
and from the lower Hubbard band to the quasiparticle band.
 The broad feature at $\omega \approx U $ at higher temperatures is  
due to transitions from the lower to the upper 
Hubbard band (compare Fig. 1).
Note that most of the spectral weight is contained
in the high frequency features.
}
\label{optcondt201}
\end{figure}

Fig. \ref{optcondt201}
shows the frequency-dependent conductivity
calculated for our model with $U = 4 t^*_1$
and $t^*_2= 0.1 t^*_1$ at three different temperatures.
It shows the important features noted below for a range of
strongly correlated metals:
(i) the Drude peak only exists at low temperatures,
and (ii) most of the spectral weight is contained in
broad high energy features.
Similar features were found previously
using dynamical mean-field theory
and exact diagonalization and iterated
perturbation theory,\cite{roz,roz2}
and for doped Mott insulators using
quantum Monte Carlo.\cite{Jarrell:95}

Infrared measurements\cite{eldridge1,eldridge2,tamura} 
of the frequency dependent conductivity $\sigma(\omega)$ of
$\kappa$-(BEDT-TTF)$_2$X 
deviate from the Drude behavior found in conventional metals.
At room temperature $\sigma(\omega)$ is
dominated by  a broad peak around
300 or 400 meV (depending on the polarization
and anion X) with a width of about 150 meV.
Even down to 50 K no Drude-like peak at
zero frequency is present (see Fig. 2 in Ref. \onlinecite{mck}).
At 25 K the high energy peak decreases slightly in
temperature and a Drude-like peak appears but
can only be fit to a Drude form if
the scattering rate and effective mass
 are frequency dependent.\cite{eldridge1}
Similar results are obtained for
$\alpha$-(BEDT-TTF)$_2$NH$_4$Hg(SCN)$_4$.\cite{dressel}

Experiments on
 $\beta$-(BEDT-TTF)$_2$X where
 X=I$_3$, IBr$_2$, and  X=AuI$_2$ at 30 K
show no Drude peak.\cite{beta}
Experiments on 
$\beta^{''}$-(BEDT-TTF)$_2$SF$_5$CH$_2$CF$_2$SO$_3$
 show no Drude peak, 
even down to 14 K.\cite{dong} 
Furthermore, it does not appear that the spectral 
weight is conserved as the temperature varies.

For (TMTSF)$_{2}$PF$_6$ at 20 K
there is a Drude peak and a broad
peak around 200 cm$^{-1}$.\cite{schwartz} The Drude peak
contains less than one per cent of the
total spectral weight and is not present
at 100 K.
 The Drude peak has been fit to a generalised
Drude form with a frequency-dependent scattering
rate $1/\tau(\omega) \sim \omega^2$,                
given by a phenomenological form used previously
for the heavy fermion compound UPt$_3$.

For SrRuO$_3$ a Drude peak was observed
at 40 K but not above about 100 K.\cite{kostic}
The conductivity $\sigma(\omega) \sim 1/\omega^{1/2}$
above a temperature dependent crossover frequency
of about $3 k_B T /\hbar$, whereas
in conventional metals, 
$\sigma(\omega) \sim 1/\omega^2$.
The low-temperature Drude peak could be
fit to a generalised Drude form with
$1/\tau(\omega) \sim \omega$.

\section{Conclusions}

In order to gain a better understanding of
why the transport properties of strongly
correlated metals deviate significantly
from the properties of elemental metals
the transport properties of a specific
Hubbard model were calculated.
The  transport properties 
are strongly temperature dependent 
because as  the temperature increases there is a smooth
crossover from coherent Fermi liquid excitations
to incoherent excitations.
This leads to a  non-monotonic temperature
dependence for the resistance, thermopower,
and Hall coefficient. The resistance smoothly
increases from a quadratic temperature dependence
at low temperatures, obeying the Kadowaki-Woods rule,
 to large values
characteristic of a bad metal.
Further signatures of the thermal destruction of
quasiparticle excitations are a peak in
the thermopower and the absence of a Drude peak 
in the optical conductivity.

The results presented here are qualitatively
similar to the observed transport properties of
a wide range of strongly correlated metals,
including transition metal oxides, strontium
ruthenates, and organic metals.    
For example,
the physical picture presents a natural explanation
of the recently presented puzzle\cite{mack2,kostic} of the properties of
SrRuO$_3$. Although Shubnikov de Haas oscillations,
with a Fermi liquid temperature dependence, were
observed at low temperatures\cite{mack2} it was found the
optical conductivity deviated significantly 
from a Drude form\cite{kostic}
and it is a bad metal at high temperatures.\cite{tyler}
This is because the latter measurements involved
energy scales (in frequency and/or temperature)
much larger than the coherence
temperature associated with Fermi liquid excitations.

Finally, it
is particularly desirable that measurements of the
temperature dependence of the thermopower
be made on a wide range of materials because
the peak that we find represents a well-defined
signature of the thermal destruction of quasiparticle
excitations. Furthermore, measurements on a single
material of {\it all} the
transport properties calculated here are needed in
order to provide a comprehensive test of the
physical picture presented.
Ideal candidate materials, since they are metallic at
ambient temperature and
have coherence temperatures of the order of 50-100 K,
are Sr$_2$RuO$_4$, $\kappa$-(BEDT-TTF)$_2$Cu(SCN)$_2$,
 and $\beta$-(BEDT-TTF)$_2$IBr$_2$.
A quantitative comparison of theory with 
experiment will require that the theory presented
here be modified to include the effects of band structure.\cite{recent}

\acknowledgements

This work was supported by the Australian 
Research Council.
We thank W. Krauth, A. Georges and M. J. Rozenberg for
helpful discussions.
Some of the computer codes used were taken from Reference 
\onlinecite{Georges:96}.

\end{document}